\documentstyle[editedvolume,numreferences,epsf]{crckapb}

\begin{opening}

\title{Irreversibility and Dephasing from Vacuum
Fluctuations} 

\author{Markus  B\"uttiker}

\institute{D\'epartement de Physique Th\'eorique, Universit\'e de Gen\`eve,\\
CH-1211 Gen\`eve 4, Switzerland}

\end{opening}
\runningtitle{Irreversibility and Dephasing}
\begin{document}
\section{Introduction}
In this work we are interested in dephasing or decoherence in the 
zero-temperature limit. The only source of decoherence 
are then provided by vacuum (zero-point) fluctuations.  
Concern with vacuum fluctuations has a long history \cite{mill}
starting with theories of black-body radiation and the Planck 
spectrum and important effects like the Lamb shift, 
the Casimir effect and the Debye-Waller factor. 
More recently, the role of vacuum fluctuations 
was discussed in theories of macroscopic quantum tunneling 
and even more closely related to our subject in theories of 
macroscopic quantum coherence \cite{legg,weiss}.  
In mesoscopic physics, we deal with systems that are so small
and are cooled to such low temperatures 
that the wave nature of electrons becomes important
and interference effects become measurable. 
Dephasing processes are therefore also of central importance 
in mesoscopic physics. Ultimately in the zero-temperature limit 
only vacuum fluctuations remain and it is 
clearly very interesting and fascinating to inquire about 
a possible role of such fluctuations. 

In contrast to the discussions on macroscopic quantum coherence,
in mesoscopic physics, the community largely 
insists that dephasing rates tend to zero (typically with some 
power law) as a function of temperature and that there is therefore
no dephasing in the zero-temperature limit.
A key argument is that in the zero-temperature limit a system 
can not excite a bath by giving away an energy quantum nor can 
a bath in the zero-temperature limit give an energy quantum 
to the system. This view holds that dephasing is necessarily 
associated with an energy transfer (a real transition) 
and since this is impossible 
there is in the zero-temperature limit no dephasing \cite{aleiner2} . 
There are, however, simple, albeit non-generic examples, in which 
decoherence is generated by the bath even so the energy 
of the small system is a constant of motion \cite{ekert}. 
However, generically, the energy of a small system coupled to a bath 
fluctuates even in the zero-temperature limit. Such fluctuations can take 
place without exciting the bath and thus without generating an energy trace 
in the bath. The fluctuations in energy of the small system are thus not 
a necessary condition for decoherence in the zero-limit, but they illustrate 
that a ground state of a system coupled to a bath is not 
as one might perhaps naively expect a quiescent state.

Many of the pioneering experiments in mesoscopic physics 
have shown a dephasing rate which saturates at low temperatures. 
The interference effect which is most often investigated results 
from the multiple elastic scattering of time-reversed electron paths
and is called {\it weak localization}.  
These experimental results are seemingly ad odds with 
theoretical predictions which find a dephasing rate 
which vanishes as the temperature tends to zero. 
Theory considers 
conductors 
in which the electrons are subject only to elastic impurity 
scattering and electron-electron interaction. Renewed attention to 
this seeming discrepancy was generated by experiments of 
Mohanty, Jariwala, and Webb \cite{mohanty1,mohanty2} 
who took extra precautions 
to avoid dephasing from unwanted sources. 
Subsequently, theoretical discussions which exhibit saturation 
were indeed presented \cite{golubev} but were strongly 
criticized \cite{aleiner1,aleiner2}. Additional sources, 
like two level fluctuators (impurities with spin,
charge traps), or some remnant radiation were
proposed in order to reach agreement with experiments.  
A detailed microscopic description of interference effects 
in a metallic conductor containing many scattering centers is impossible. 
Instead, the procedure is to investigate only the average 
over an ensemble of metallic diffusive conductors with each 
ensemble member having a different impurity configuration. 
After ensemble averaging the conductor becomes translationally 
invariant and it is this fact which permits to make analytical progress. 
The fact that in weak localization we deal not only with 
quantum and statistical mechanical averages 
but also with disorder averages makes its discussion heavy 
and let's us search for simpler model systems in which these
conceptual issues can be discussed. 
We can sum up this discussion by quoting a 
sentence from a recent publication. 
Ref. \cite{natel} states 
"An intrinsic saturation of $\tau_\phi$ (the dephasing time) 
to a finite value as $T \rightarrow 0 $ would have profound implications 
for the ground state 
of metals and might indicate a fundamental limitation in controlling 
quantum coherence in conductors".

The mesoscopic systems considered here consists of a ring 
penetrated by an Aharonov-Bohm flux. If electron-motion is coherent 
such a system exhibits an interesting ground state:  
there exists a persistent current. We will further assume 
that the ring contains an in-line quantum dot which is weakly coupled to the 
arms of the ring. A persistent current exists then only 
if the free energy of the system with $N$ electrons on the dot 
is (nearly) equal to the free energy with $N+1$ 
electrons on the dot. 
We have thus an effective two level problem \cite{mbcs}: 
when coupled to a 
bath (a resistor capacitively coupled to the dot and the arms of the ring) 
the model is equivalent to a spin-boson problem 
\cite{cedr}. 
The spin boson 
problem is a widely discussed system with a ground state 
that depends on the coupling strength to the bath \cite{legg,weiss,chak,bray}. 

Are there any fundamental requirements which must be present 
in order to have decoherence? Since energy exchange is not necessary 
one might wonder what the essential ingredient in a dephasing process
is. It is sometimes stated that entanglement is crucial, that is the fact 
that the state of the system and bath is not a product state. Entanglement 
is necessary but it 
is a natural and generic byproduct of the coupling of any two 
systems \cite{loss}. As long as the combined system is governed 
by a reversible time evolution, entanglement alone can not 
generate dephasing. In this work we regard the {\it irreversibility}
provided by the bath as the essential source of decoherence.
If the state of the bath, evolves in time in a way that it never returns,
then if we regard the system alone, typically, 
its states will dephase in time.

Vacuum fluctuations and their interaction with a small system 
can provide the necessary source of irreversibility. 
To illustrate this we first discuss the simple case of 
an $LC$-oscillator coupled to a transmission line. 
This can be formulated as a scattering problem with 
an incident vacuum field and an outgoing vacuum field. 
We discuss the zero-temperature fluctuations of the energy 
of the $LC$-oscillator and investigate the trace which the oscillator 
leaves in the bath. 
We next consider systems which effectively can be mapped 
onto two-level systems. A two-level system for which the 
Hamiltonian of the isolated system commutes with the Hamiltonian of the 
total system provides a simple, exactly solvable example, 
in which excited states decoher but for which the ground state is 
a pure state. We next consider effective two level systems 
for which the Hamiltonian of the system does not commute with 
the total Hamiltonian (system + bath). As a special example 
we discuss a mesoscopic system which exhibits a persistent current
in its ground state and discuss the suppression of the persistent 
current and the increase in its fluctuations 
with increasing coupling to the bath. This is an example 
of a system in which vacuum fluctuations generate a 
partially coherent ground state.

\begin{figure}[ht]
\vspace*{0.5cm}
\centerline{\epsfysize=5cm\epsfbox{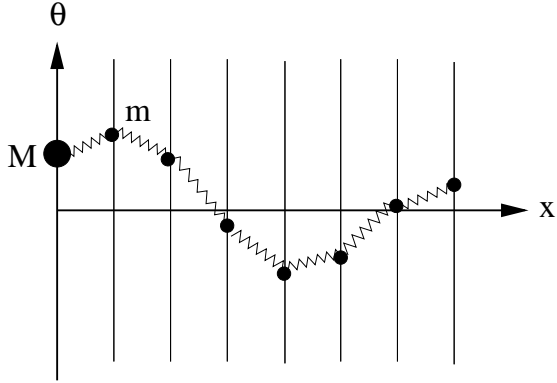}}
\vspace*{0.5cm}
\caption{\label{pchain}
A particle at $x= 0$ with mass M moving in a potential $V(\Theta)$
is coupled to particles at $x = na$, $n = 1, 2, 3, ..$ forming 
an infinite string.  
}
\end{figure}

\section{The Lamb model} 
\label{lamb} 

The model which we consider is shown in Fig. \ref{pchain}. 
A particle with mass $M$ moves at $x = 0$ along 
the $\Theta$-axis in a potential $V(\Theta)$. 
It is coupled to a harmonic chain with particles at 
positions $x = na$ and with elongation 
$\Theta_{n}$. This model was 
introduced by Lamb \cite{lamb} in 1900 with the purpose of
understanding
the then new notion of radiation reaction in electrodynamics.
Any movement of the particle at $x = 0$ (the "system") 
generates a wave in the harmonic chain that travels 
to the right away to infinity.
Waves on the harmonic chain can be written as a superposition 
and can be separated into left moving traveling waves $\Theta_{L} (x/v + t)$
and right moving traveling waves $\Theta_{R} (x/v - t)$.
The incident waves (the right movers) come from infinity 
and their properties are determined by an equilibrium 
statistical ensemble. These waves are incident on the system where they 
are reflected and propagate back to infinity.  
This model of a bath makes immediately apparent 
how a purely Hamiltonian bath generates irreversibility. 
It is "the traveling away to infinity"
which is the source of irreversibility. 
In turn, this irreversible behavior, is the key ingredient which 
distroys  the coherence of the "system".  

The Lamb model is equivalent to a number of models using harmonic 
oscillators to describe a bath \cite{ford}. Which of these models one uses 
is a matter of taste. In addition to the physically transparent way 
in which irreversibility is generated in the Lamb model, 
one more reason why we prefer it here 
is that it permits a scattering approach
to describe the interaction between the system and the bath.

\begin{figure}[ht]
\vspace*{0.5cm}
\centerline{\epsfysize=2.5cm\epsfbox{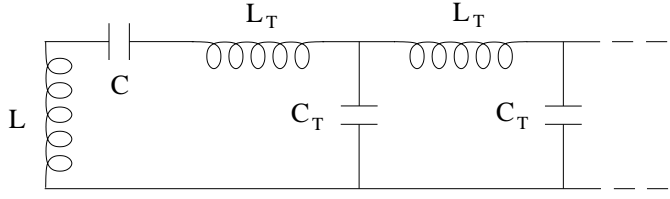}}
\vspace*{0.5cm}
\caption{\label{tline}
$LC$-oscillator coupled to a transmission line with 
inductances $L_T$ and capacitances $C_T$. For the 
$LC$-oscillator 
the transmission line acts like a resistor with resistance 
$R = (L_{T}/C_{T})^{1/2}$. 
}
\end{figure}

Yurke and Denker \cite{yude} 
pioneered a description of such models in terms of a scattering problem 
for incoming and outgoing field states. Following them 
we consider the system shown in Fig. \ref{tline}. 
It is the electrical circuit equivalent to the 
mechanical system shown in Fig. \ref{pchain}.
We assume that the "system" is a simple "LC"-oscillator
with an inductance $L$ and a capacitance $C$. 
This system is coupled to an external circuit 
which is represented by a transmission line 
with inductances $L_{T}$ and capacitances $C_{T}$. 

In the absence of the transmission line, the $LC$-oscillator 
has an energy
\begin{eqnarray}
H_{0} = (L/2)  \, I^{2} + (1/2C) \, Q^{2} .
\label{hs}
\end{eqnarray}
To couple the single oscillator to the transmission 
line we sum over all charges 
to the right of the $n-1-th$ capacitor, 
$Q(na,t) = \sum_{m = n}^{\infty} Q_{m}$
and consider the continuum limit, $Q(x,t)$. 
The current along the transmission line is 
$I(x,t) = \partial Q(x,t)/\partial t$, 
and the voltage across the transmission line 
$V(x,t) = - c_{T}^{-1}\partial Q(x,t)/\partial x$.
Here $c_{T}$ is the capacitance per unit length and 
$l_T$ is the inductance per unit length. 
We separate the charge field in left and right traveling 
waves $Q_{L}(x/v+t)$ and $Q_{R}(x/v+t)$
with velocity $v = (l_{T}c_{T})^{-1}$.
The equations of motion of the combined system are 
\begin{eqnarray}
L \, d^{2}Q/dt^{2} + R \, dQ/dt + Q/C = 2R \, dQ_{L}/dt 
\label{eqm0}
\end{eqnarray}
and for $x > 0$ 
\begin{eqnarray}
d^{2}Q/dt^{2} - v ^{2} \, dQ/dx^{2}dt = 0. 
\label{eqm1}
\end{eqnarray}
The charge on the capacitor of the oscillator is 
$Q(t) \equiv Q(0,t)$. Due the transmission line the 
oscillator "sees" effectively an external circuit 
with a resistance $R = (l_{T}/c_{T})^{1/2}$.  
It is important to note that in Eq. (\ref{eqm0}) the noise term is 
described by the incoming field (left movers). The out-going field 
(right movers) is completely determined by the boundary condition 
$Q(t,x) = Q_{L} (x/v+t) + Q_{R} (x/v-t)$ at $x = 0$, since 
the incident field is known, and 
the charge $Q(t,x =0)$ is determined by Eq. (\ref{eqm0}).  

It is easy to quantize this system \cite{yude,gard}. For instance 
the incoming field is described by 
\begin{eqnarray}
\label{field} 
Q_{L}(x/v+t) =  (\hbar /4\pi R)^{1/2} \int_{0}^{\infty}  
d\omega \omega^{-1/2} \nonumber\\ 
                \left(a_{L\omega} \, \exp(-i\omega(x/v+t)) 
                + a_{L\omega}^{\dagger} \, \exp(i\omega(x/v+t))\right)              
\end{eqnarray} 
where the $a_{L\omega}$ are bosonic annihilation operators. 
Similarly we can write an expression for the right moving field 
in terms of annihilation operators $a_{R\omega}$. 
If we denote the annihilation operator for the oscillator at $x = 0$ 
(the "system") by $a_{\omega}$ then the relation 
$Q(t,x) = Q_{L} (t,x) + Q_{R} (t,x)$ at $x = 0$
implies that $a_{\omega} = a_{L\omega} + a_{R\omega}$.
Using this and observing that Eq. (\ref{eqm0}) connects 
$a_{\omega}$ and $a_{L\omega}$ via 
\begin{eqnarray}
(\omega^{2} - \omega^{2}_{0} + i\eta \omega ) \, a_{\omega} = 
2i \eta \omega \, a_{L\omega}
\end{eqnarray} 
gives \cite{yude} 
\begin{eqnarray}
a_{R\omega} = s(\omega) \, a_{L\omega}
\end{eqnarray}
with a {\it scattering matrix} 
\begin{eqnarray} 
s(\omega) = - \frac{\omega^{2} - \omega^{2}_{0} - i\eta \omega}
{\omega^{2} - \omega^{2}_{0} + i \eta \omega} .
\label{sma}
\end{eqnarray}
Here we have introduced the oscillation frequency of the 
decoupled oscillator $\omega_{0} = (LC)^{-1}$
and the friction constant $\eta = R/L$. 

All incoming field modes are ultimately reflected.
Thus the reflection probability is $|s(\omega)|^{2} = 1$. 
The incoming field modes are scattered elastically and the only 
effect of the scattering process is a phase-shift 
$s(\omega) = - e^{i\phi(\omega)}$. 
In the weak damping limit the phase-delay time is 
$d\phi/d\omega = \eta/[(\omega^{2} - \omega^{2}_{0})^{2} + 
\eta^{2} /4]$. It peaks for modes incident 
at the frequency $\omega_{0}$ of the oscillator. 
This leads to a picture in which the energy loss 
due to radiation damping of 
the oscillator is compensated by the energy supplied 
by the incoming vacuum fluctuations. 
The mean energy of the oscillator is simply determined by the 
balance of these two effects, whereas the fluctuations in energy 
result from the fact that the energy supply 
from the vacuum is itself a fluctuating process. 
With the incident field as specified above, the commutation 
rules and the fact that in the zero-temperature limit 
any annihilation operator 
of the incoming field applied to the vacuum gives zero,  
$a_{L\omega}|0> = 0$, all the quantities of interest can be calculated.  
Next we simply state a few results of this model.

\section{Fluctuations in the ground state of the Lamb model} 

\begin{figure}[ht]
\vspace*{0.5cm}
\centerline{\epsfysize=6cm\epsfbox{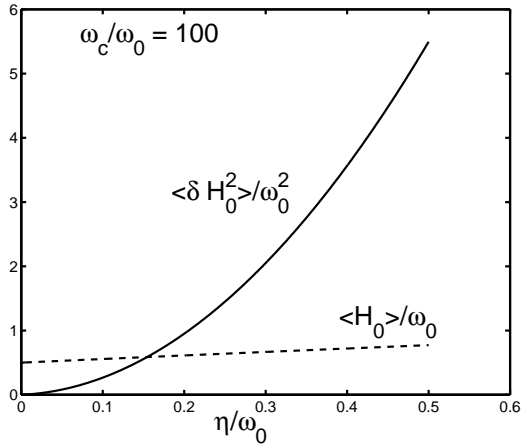}}
\vspace*{0.5cm}
\caption{\label{hfluct}
The mean energy $<H_{0}>$ (broken line) 
of the oscillator and the instantaneous mean-squared energy fluctuations 
$<\delta H^{2}_{0}> = < (H_{0}(t)- <H_{0}>)^{2}>$
in the ground state of the Lamb model as a function of the
coupling constant $\eta$ in units with $\hbar = 1$. 
After K. Nagaev and M. B\protect\"uttiker \protect\cite{naga}.  
}
\end{figure}

Of interest here are the properties of the ground state of the Lamb model. 
Since the model is exactly solvable the results presented below 
can be given for arbitrary coupling strength $\eta$. However, 
in order to be brief we focus here on the low-coupling limit
and give the results only to leading order in $\eta$. 
The results for the mean-squared fluctuations of the 
charge 
and for the mean-squared current are frequently quoted \cite{weiss}. 
To leading order in $\eta$ the mean-squared fluctuation 
of the charge are reduced below their value in the uncoupled 
system, 
\begin{eqnarray}
\frac{<Q^{2}>}{C} = \frac{\hbar \omega_{0}}{2} \, 
\left( 1 - \frac{\eta}{\pi{\omega_{0}}}\right) .
\end{eqnarray}
On the other hand, coupling to the bath increases the 
momentum fluctuations. 
In fact without a high frequency cut-off $\omega_{c}$ 
these fluctuations would diverge. We have  
\begin{eqnarray}
L \, <I^{2}> = L \omega_{0}^{2} \, <Q^{2}> + \frac{\hbar \eta }{\pi} \, 
\ln\left(\frac{\omega_{c}}{\omega_{0}}\right) .  
\end{eqnarray}
From these results we immediately obtain the mean energy 
of the oscillator. We see that 
the mean energy of the oscillator is not simply 
$\hbar \omega_{0}/2$ but like the momentum would diverge 
without taking into account a cut-off. 
To leading order in $\eta$ we have instead 
\begin{eqnarray}
<0|H_{0}(t) |0> = \frac{\hbar {\omega_{0}}}{2} + 
(\frac{\hbar \eta}{2\pi}) \, [ \ln\left(\frac{\omega_{c}}{\omega_{0}}\right) -1]
\label{eaverage} 
\end{eqnarray} 
The expectation value of the energy is shown in Fig. \ref{hfluct} 
as a function of $\eta$ (over a range of $\eta$ exceeding 
the validity of Eq. (\ref{eaverage})). 

In the uncoupled system the energy of the oscillator does not 
fluctuate. On the other hand, for the oscillator coupled 
to the bath, its energy needs not to be well defined. Energy conservation 
applies only to the total system but not to a subsystem \cite{cedr}. 
Consider the energy operator $H_{0}$ given by Eq. (\ref{hs}).   
A calculation by Nagaev 
and the author \cite{naga} shows indeed that the expectation value 
of $<0|(\Delta H_{0}(t))^{2}|0>$
where $\Delta H_{0}(t) = H_{0}(t) - <0|H_{0}(t) |0>$ 
is given by 
\begin{eqnarray}
<0|(\Delta H_{0}(t))^{2}|0> = \frac{\hbar^{2} \omega_{0} \eta}{2\pi} \, 
[\ln(\frac{\omega_{c}}{\omega_{0}}) - 1] 
\label{efluct}
\end{eqnarray}
In the weak coupling limit the energy fluctuations are proportional 
to the average excess energy $(<0|H_{0}(t) |0> - \hbar \omega_{0}/2)$. 
The factor of proportionality is 
$\hbar \omega_{0}/2$.  

The mean-squared fluctuations of the energy of the oscillator 
are shown in Fig. 3 as a function of $\eta$
(over a range exceeding the limit of validity of Eq. (\ref{efluct})). 
We have emphasized the possibility of energy fluctuations
in the ground state, since following Ref. \cite{cedr}, 
the possibility of such fluctuations was questioned in Ref. \cite{gavish}. 

It is perhaps necessary to point out the following: 
In standard statistical mechanics it is assumed that the coupling energy 
is always smaller than any relevant energy scale of the system under 
consideration \cite{becker}.  
Thus in standard statistical mechanics
the only properties of the bath which enter are the temperature 
and the chemical potential of the bath. Here in contrast, 
the friction constant $\eta$ enters in a non-trivial way. 
We next aim at characterizing the vacuum state of the bath.

\section{Traces in the bath} 
\label{trace} 

In theories of decoherence \cite{zure,sia} the trace which a system leaves 
in the environment plays an important role. 
Zurek compares this to the waves a ship leaves in its wake 
while crossing a sea \cite{zure}. Since the vacuum 
is not a quiet state, we can ask: 
Does the oscillator leave a trace in the bath even in the ground state of 
the system? And if the answer is yes, what are the properties of 
this trace? 

First, let us characterize the incoming field. 
The noise force 
$2 V_{L} (t) = - 2 R dQ_{L} (t)/dt $ (see Eq. (\ref{eqm0})) 
which acts on the harmonic oscillator is 
determined by the voltage drop across the oscillator generated by 
the incident (left moving) field. Thus we will discuss the 
state of the bath by investigating the voltages generated 
by the left and right moving traveling fields. 

A simple calculation gives 
for the correlation $S_{V_{L} V_{L}} (t) \equiv \langle 0|V_{L} (t) V_{L} (0) 
+ V_{L} (0) V_{L} (t)|0 \rangle $ 
\begin{eqnarray}
\label{vtrav} 
S_{V_{L} V_{L}} (t) = 
\frac{\hbar R}{4\pi} \, \int_{-\infty}^{\infty} d\omega
|\omega| \, \exp(-i\omega t) \exp(-\omega/\omega_{c}) = \nonumber\\
\frac{1}{e^{2}} \, \frac{R}{R_K} \, \frac{(\hbar \omega_{c})^{2}}{2\pi}
\frac{1 - \omega_{c}^{2} t^{2}}{(1 + \omega_{c}^{2} t^{2})^{2}} .
\end{eqnarray}
Here we have introduced the resistance quantum $R_K = h/e^{2}$. 
The correlation is positive at very short times $t < 1/\omega_{c}$
but negative at long times with a decay
proportional to  $t^{-2}$. 
This property is a consequence of the particular frequency 
dependence of the noise power spectral density of the vacuum fluctuations.
The spectral density is 
\begin{eqnarray}
<0|V_{L} (\omega) V_{L} (\omega^{\prime}) + 
V_{L}(\omega^{\prime}) V_{L} (\omega)|0>
= 2\pi S_{V_{L}V_{L}}(\omega) \, \delta(\omega + \omega^{\prime})
\end{eqnarray}
where $V(\omega)$ is the Fourier transform of $V(t)$. 
The vacuum fluctuations of the force are 
characterized by a spectral density 
\begin{eqnarray}
S_{V_{L}V_{L}}(\omega) = \hbar |\omega| \, R /4  
\label{svac} 
\end{eqnarray}
Now since the incoming field $V_{L} (x/v+t)$ differs from the 
field at $x = 0$ only by a phase factor multiplying each 
operator,  
Eq. (\ref{svac})
gives also the spectral density of the voltage correlation
$<0|V_{L} (x,t) V_{L} (x,0) + V_{L} (x,0) V_{L} (x,t)|0>$. 
at any other 
point $x$ along the chain. 
Next consider the voltage $V_{R} (x/v-t)$ generated by the right movers. 
Its spectral density is also given by Eq. (\ref{svac}). 
Both the left and right goers 
are characterized by the same spectral density. 

To find the trace which the system leaves in the environment 
we have to consider the correlations 
between the voltage fluctuations generated by the incident field
and the voltage fluctuations of the out-going field. 
We find that these correlations have a spectral density 
\begin{eqnarray}
S_{V_{L}V_{R}} (\omega) = S_{V_{R}V_{L}} (- \omega) = 
(\hbar Z/4) \, |\omega| \,  s(\omega) 
\end{eqnarray}
where $s$ is the scattering matrix given by Eq. (\ref{sma}).

The spectrum of the total voltage fluctuations
(which contains the correlations between left and right movers) 
exhibits a dramatic effect. 
From the spectral densities of the incoming voltage fluctuations 
and the outgoing voltage fluctuations we can calculate 
the spectrum of the total voltage fluctuations, $V = V_{L} + V_{R}$
for which we find,  

\begin{eqnarray}
S_{VV} (\omega) =
(\hbar |\omega|R/2) \, (2 + s(\omega) + s(-\omega)) 
\end{eqnarray}
which with the $s$-matrix, Eq. (\ref{sma}) gives
\begin{eqnarray}
S_{VV} (\omega) = R \,  \hbar|\omega| \, 
\frac{(\omega^{2} - \omega^{2}_{0})^{2}} 
{(\omega^{2} - \omega^{2}_{0})^{2} + \eta^{2} \omega^{2}} .
\label{vol}
\end{eqnarray}
At resonance, for $\omega = \pm \omega_{0}$, 
the spectral density of the total voltage fluctuations
vanishes. This reflects the fact that at resonance our system can not 
maintain a voltage drop.  

Eq. (\ref{vol}) can be simply derived 
from an equivalent electrical circuit. 
We replace the transmission line 
by a resistor with  
resistance $R$ 
in parallel with a current noise source with 
current $I_{N}$.  The current noise source 
has a spectral density 
\begin{eqnarray}
S_{I_{N}I_{N}} (\omega) = (1/R) \, \hbar|\omega| \, coth(\hbar \omega /2kT) .
\label{noises}
\end{eqnarray}
Here we have for later reference given the finite temperature result. 
Unless stated otherwise we continue to discuss 
the zero-temperature limit only. 
The current through the "system" (the LC-oscillator) is 
$I_{s} = V/Z_{s}$. Here  $Z_{s} = (-i \omega C)^{-1} -i \omega L$
is the impedance of the system as seen from the noise source. 
The current through the resistance is $I = V/R$. 
Current conservation demands $I_{s} + I + I_{N} = 0 $ and 
together with Eq. (\ref{noises}) leads immediately to 
the spectral density Eq. (\ref{vol}) for the voltage $V$. 

Thus far we have considered a harmonic oscillator at zero-temperature 
coupled to a bath. We have shown that its energy fluctuates and have 
examined the trace it leaves in the bath. This trace has its source in the 
time-delay suffered by vacuum fluctuations due to their 
interaction with system. 
Next we now consider two level systems coupled to a bath 
and consider dephasing and and decoherence in these systems.

\section{Two-Level Systems}

\begin{figure}[ht]
\vspace*{0.5cm}
\centerline{\epsfysize=6cm\epsfbox{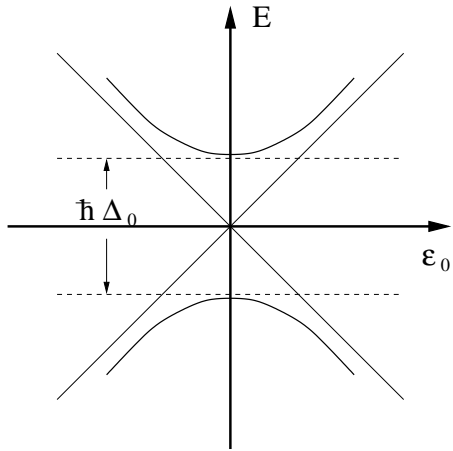}}
\vspace*{0.5cm}
\caption{\label{twolfig}
Energy levels of the two state system in the presence of 
tunneling ($\Delta_{0} \ne 0$) and in the absence of 
tunneling ($\Delta_{0} = 0$).
}
\end{figure}

A large class of physically interesting systems can be reduced to an
effective two-level
problem. We consider systems which in the absence of a bath are described
by the 
Hamiltonian
\begin{equation}
\label{ham2} 
H_{0} = \frac{\hbar \epsilon_{0}}{2} \, \sigma_{z} - 
\frac{\hbar \Delta_{0}}{2} \, \sigma_{x} 
\end{equation}
where $\hbar \epsilon_{0}$ is the energy difference between two levels 
(in the absence of tunneling), 
and $\Delta_{0}$ is a tunneling frequency which hybridizes the two
levels. The energy of the ground state and the excited state of this
system are 
\begin{equation}
E_{\mp} = \mp \frac{\hbar \Omega_{0}}{2} 
= \mp \frac{\hbar}{2} \sqrt{\epsilon_{0}^{2}+ \Delta_{0}^{2}} 
\end{equation}
where we have introduced the frequency $\Omega_{0}$. 
The two energy levels are shown in Fig. \ref{twolfig} 
as a function of $\epsilon_{0}$. 
The coupling energy is
\begin{equation}
H_{c} = g \frac{eV}{2} \, \sigma_{z}
\end{equation}
where $g$ is a dimension less coupling constant and $V$ is the voltage
which drops 
across the system. The total Hamiltonian contains in addition 
the contribution from all the energies of the bath oscillators.

\begin{figure}[ht]
\centerline{\epsfysize=6cm\epsfbox{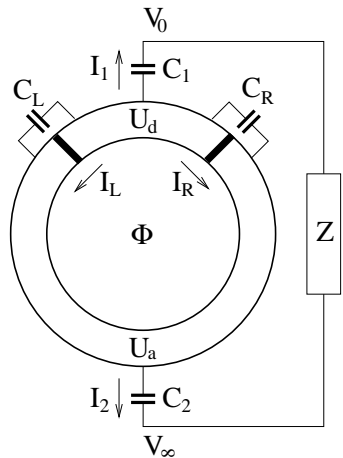}}
\vspace*{0.3cm}
\caption{\label{system} Ring with an in-line dot subject to a 
flux $\Phi$ and capacitively coupled to an external impedance $Z$.
After \protect\cite{cedr,lange}.
}
\end{figure}

To be specific we consider the electrical system 
shown in Fig. \ref{system}. 
A small mesoscopic ring is formed by a quantum dot 
which is coupled via tunnel barriers with transmission amplitudes
$t_{L}$ and $t_{R}$ to leads.
The leads are connected back onto themselves such that they form 
together with the dot a ring structure.
An Aharonov-Bohm flux $\Phi$ penetrates the hole of the ring.
The model without the external circuit is discussed in 
Ref. \cite{mbcs}. The role of the external circuit 
(the bath) is the subject of Refs. \cite{cedr,annals,lange}. 
A ballistic ring coupled to the electromagentic vacuum fields 
is investigated in Ref. \cite{loma}. In our geometry, the quantum dot 
generates a system that is highly sensitive to fluctuations 
of the environment. 
A completely equivalent model consists of a 
superconducting electron box \cite{bouch} 
which can be opened to admit an Aharonov-Bohm flux \cite{moji,makh}.

The arm of the ring contains electrons in 
levels with energy $E_{am}$ and the dot contains electrons 
in levels with energy $E_{dn}$. First let us for a moment
neglect tunneling. Let $F_{N}$ be the free energy 
for the case that there are $N$ electrons in the dot and $M$
electrons in the arm. The transfer of an electron from the 
arm to the dot gives a free energy $F_{N+1}$.
The difference of these two free energies 
is
\begin{equation}
\label{detune}
\hbar \epsilon_{0} = F_{N+1}-F_{N} = E_{d(N+1)}-E_{aM} + 
\frac{e^{2} (N + 1/2 - C_{0}V_{e})}{C} . 
\end{equation}
Here the first two terms arise from the difference in kinetic energies.
The third term results from the charging energy of the dot.
$C^{-1}_{0} = C^{-1}_{i} + C^{-1}_{e}$ is the series capacitance
of the internal capacitance $C_{i} = C_{L} + C_{R}$ 
and the external capacitance $C^{-1}_{e} = C^{-1}_{1} + C^{-1}_{2}$.
The total capacitance is  $C = C_{i} + C_{e}$. 
Eq. (\ref{detune})
measures the distance to resonance, 
$\epsilon_{0} = 0$ is the condition that the 
Coulomb blockade is lifted \cite{bee}. 

Tunneling through the barriers connecting the dot and 
the arm is described by amplitudes $t_{L}$ and $t_{R}$. 
The total tunneling energy through the dot depends on these
amplitudes and the Aharonov-Bohm flux $\Phi$ in the following way, 
\begin{equation}
\label{Delta0}
{\hbar \Delta_0 \over 2}
= \sqrt{t_L^2 + t_R^2 \pm 2 t_L t_R \cos{2\pi \Phi \over \Phi_0}},
\end{equation}
where $\Phi_0 = hc/e$ is the single electron flux quantum. 
The sign depends on the number of electrons in the 
dot and arm: it is positive, the total number is odd, and it is 
negative if the total number is even \cite{mbcs}. 
The voltage across the system is $V = V_{0} - V_{\infty}$. 
The coupling constant in $H_{c}$ is $g = C_{0}/C_{i}$. 

In the two-level limit of interest here the transmission 
amplitudes $t_L$ and $t_R$ are taken to be very small 
compared to the level spacing in the dot and in the arm. 
For larger transmission amplitudes the system without 
a bath will already exhibit a Kondo effect \cite{eckle,kang,affl,hu}.

There are two important classes of two-level systems, 
depending on whether 
$\sigma_{z}$ commutes with the total Hamiltonian or not.

\section{Two-level systems with a coherent ground state}

Consider the case of vanishing tunneling frequency $\Delta_{0} = 0$.
In this case $[H,\sigma_{z}] = 0$. As a consequence a pure state 
state of $H_{0}$ remains a {\it pure} state also in the presence 
of the coupling to the bath. 

This model is most often 
used to investigate the decoherence of superpositions 
of {\it two} states. We discuss 
this briefly even though our main interest 
is in the coherence properties
of the ground state and not of the excited states 
of the system. The state of the system can be written 
as a spinor with angles $\theta$ and $\phi$ on the Bloch 
sphere,
\begin{equation}
    |\Psi >  =  \left( \begin{array}{ll}
                \cos(\theta/2) \, \exp(i\phi/2)\\
                \sin(\theta/2) \, \exp(-i\phi/2)
                \end{array} \right)
\label{spinor} .
\end{equation}
This state represents a superposition of a spin up state
$|\uparrow > = 
{0 \hfill \choose 1},$ and a spin 
down state
$|\downarrow > = 
{1 \hfill \choose 0},$ with
\begin{equation}
|\psi > = \cos(\theta/2) \, \exp(i\phi/2)| \uparrow > + 
\sin(\theta/2) \, \exp(-i\phi/2) | \downarrow >. 
\end{equation}
In the absence of a bath the spin precesses with a Larmor frequency
$\omega_{L} = \epsilon_{0}$ around the $z$-axis, 
such that $\theta = const$ and 
$\phi = - \omega_{L}t$.
If the system is coupled to the bath it is still possible to find 
an exact solution of the full problem for arbitrary coupling strength.
Palma, Souminen and Ekert \cite{ekert} discuss this in terms of displaced
bath oscillators.   
Here, to find this solution 
we proceed by looking at the problem from an electrical circuit point of
view and derive a simple Langevin equation. 
As briefly discussed in Section \ref{trace}, 
we can represent the bath by a
resistor $R$ in parallel with 
a noise source with current $I_{N}(t)$ 
with a spectral density given by Eq. (\ref{noises}). 

Using the spinor Eq. (\ref{spinor})
with $\phi (t) = - \omega_{L} t + \delta \phi (t)$, where $\delta \phi(t)$
accounts for the fluctuations in the phase, we find from the Hamiltonian 
the equation 
\begin{equation}
\frac{d\delta \phi (t)}{dt} = \frac{e}{\hbar} \,  \frac{C_{0} }{C_{i}} \, 
V(t) 
\label{two1}
\end{equation}
Next we need to find the voltage 
which drops across this two-level system.
The current through the resistor is $V/R$, the
current through the 
system is $C_{0} dV(t)/dt$ and the current of the noise source 
in parallel to the resistor is $I_{N}(t)$ and therefore,
\begin{equation}
V = - RC_{0} \,  dV
(t)/dt +  R \, I_{N}(t) . 
\label{two2}
\end{equation}
Scattering of the incident vacuum states at this system 
is described by an s-matrix $s = - (1 + i\omega \tau)/(1 - i\omega \tau)$
which is the overdamped limit $\omega^{2} \ll \omega_{0}^{2}$ 
of Eq. (\ref{sma}). Here $\tau = \eta / \omega_{0}^{2} = RC_{0}$ 
is the $RC$-time of the electrical circuit. 
Taking the Fourier transform of Eqs. (\ref{two1}) and  (\ref{two2}) 
gives for the 
spectral density of the voltage fluctuations
\begin{equation}
S_{VV}(\omega) = \frac{R \hbar |\omega|}
{1 + \omega^{2} \tau^{2}}
\label{vspect1} 
\end{equation}
and for the phase fluctuations 
\begin{equation}
S_{\phi \phi}(\omega) = 
\frac{e^{2}}{\hbar^{2}} \, \frac{C_{0}^{2} }{C_{i}^{2}} \, 
\frac{1}{\omega^{2}} \, 
\frac{R \hbar |\omega|} 
{1 + \omega^{2} \tau^{2}} . 
\label{pspect} 
\end{equation}
The spectrum is proportional to $1/|\omega |$ 
at low frequencies. 
Consider next the density matrix. 
We have 
\begin{equation} \label{sm}
{\bf \rho} = 
\left( \matrix{ \cos^{2}(\theta/2) & (1/2) \sin(\theta) \, \exp(i\phi)\cr 
(1/2) \sin(\theta) \, \exp(-i\phi)& \sin^{2}(\theta/2)} \right).
\end{equation} 
Since the fluctuations $\delta \phi (t)$ are Gaussian 
(as is known for a harmonic oscillator coupled to a bath) 
we find that the averaged density matrix evolves away from 
its initial value at $t = 0$ according to 
$<\exp(i\phi) > =$ $\exp(-i \omega_{L} t)$ $\exp(-\Gamma (t))$
where 
\begin{equation}
\Gamma (t) = (1/2) <(\phi (t) - \phi (0))^{2}> . 
\label{gamma} 
\end{equation}
With the help of the spectral density Eq. (\ref{pspect}) we find the 
mean-square deviation of the phase away from its initial 
value, 
\begin{equation}
\Gamma (t) = 
(1/\pi) \int d \omega \, S_{\phi \phi}(\omega) \, \sin^{2}(\omega t/2) . 
\label{phiev} 
\end{equation}
Thus $\Gamma (t)$ determines the decoherence 
of a superposition of the two states of the two-level system. 
Our spectrum Eq. (\ref{vspect1}) has a natural cut-off due to 
lorentzian role-off of the spectrum with 
the $RC_{0}$-time. This gives $\Gamma (t) \propto (t/\tau)^{2}$
for $t \ll \tau$ and gives $\Gamma (t) \propto \ln (t/\tau)^{2}$
at long times, and thus an approximate expression covering both 
the short and long time behavior is 
\begin{equation}
\Gamma (t) \approx 2 \alpha \, 
\ln\left(1+ (t/4\tau)^{2}\right) . 
\end{equation}
where we have introduced the coupling parameter 
\begin{equation}
\label{coupp}
\alpha \equiv {R \over R_K} \, \left( {C_0 \over C_i} \right)^2 . 
\end{equation}
Here $R_K = h/e^{2}$ is the von Klitzing resistance quantum. 
For $t > \tau $ the decay of the two state superposition 
is $\exp(-\Gamma (t)) \propto t^{-2\alpha}$. 

To summarize: the two-level system with 
$\Delta_{0} = 0$ is a simple example with a coherent ground state. 
Superpositions of the ground state and the excited state decay 
due to the vacuum fluctuations of the bath. This model exhibits no 
population decay (the diagonal elements of the density matrix
are constants of motion). Initial off-diagonal elements of the density 
matrix vanish over time due vacuum fluctuations. 
Next we consider the case of a two-level system 
with a tunneling frequency given by $\Delta_{0}$.

\section{Two-level systems with a partially coherent ground state}

In the presence of an Aharonov-Bohm flux the ground state of 
the ring-dot system shown in Fig. \ref{system}
permits a persistent current if the two tunneling amplitudes 
$t_{L}$ and $t_{R}$ are non-vanishing. If we consider for a 
moment the system decoupled from the bath, the free energy ( apart
from an unimportant constant) is 
\begin{equation}
\label{freee} 
\Delta F = - 
\frac{\hbar}{2}  \sqrt{\epsilon_{0}^{2} + \Delta_{0}^{2}}
\equiv  - \frac{\hbar \Omega_{0}}{2} . 
\end{equation}
The second equality defines the frequency $\Omega_{0}$.
Its derivative with respect to an Aharonov-Bohm flux $\Phi$
gives an equilibrium, ground state current, 
given by \cite{mbcs}  
\begin{equation}
<I> = -c \frac{d\Delta F}{d\Phi} = \mp \, {e}
\frac{4 \pi t_{L} t_{R}} {\Omega_{0}} \, 
\sin(2\pi \Phi/\Phi_{0}). 
\end{equation}
The equilibrium current is a pure quantum effect: only 
electrons whose wave functions are sufficiently 
coherent to reach around the loop contribute 
to the persistent current. Thus the persistent current is
a measure of the coherence of the ground state. 
At resonance $\epsilon = 0$ the current is of the 
order of $et$ with $t$ a transmission amplitude and 
it decreases and becomes of the order 
of $et^{2}/\epsilon $ as we move away from resonance.  

\begin{figure}[ht]
\centerline{\epsfysize6.5cm\epsfbox{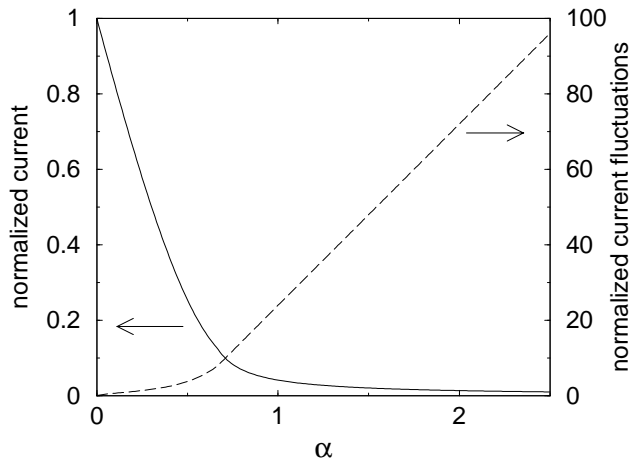}}
\caption{\label{kondopc} The persistent current $<I>$ at resonance
$\epsilon = 0$ for a symmetric ring (solid line) 
and the fluctuations of the
circulating current $(<(I (t) -  <I (t)>)^{2}>)^{1/2}/<I>$
(dashed line)
as a function of the coupling parameter $\alpha$.   
The average persistent current is in units of the persistent 
current at $\alpha = 0$. The fluctuations are normalized by the average 
current. The parameters are $\omega_c = 25 \Delta_0$.
After \protect\cite{cedr}.}
\end{figure}

Therefore, it is interesting to ask how this current is affected 
if the system is coupled to a bath. 
In contrast to the two-level problem in the absence of 
tunneling, the two-level problem of interest here, 
is not exactly solvable.  
Cedraschi, Ponomarenko and the author \cite{cedr,annals} used 
known solutions from a Bethe ansatz 
and perturbation theory 
to provide an answer. In addition these authors also 
investigated the (instantaneous) fluctuations 
of the equilibrium current away from its average, 
$<(I (t) -  <I (t)>)^{2}>$. 
For the symmetric case $t_{L} = t_{R}$, 
$C_{L} = C_{R}$,  the average current 
together with the mean-squared current fluctuations 
is shown in Fig. \ref{kondopc} as a function 
of the coupling parameter $\alpha$ (see Eq. (\ref{coupp})). 
The persistent current 
is in units of the current for $\alpha = 0$. 
A similar, but very weak, reduction 
of the persistent current
has also been found for a purely ballistic ring 
coupled to the electromagnetic vacuum fields \cite{loma}. 
In Fig. \ref{kondopc}
the mean squared current fluctuations $(<(I (t) -  <I (t)>)^{2}>)^{1/2}$
are in units of the average current $<I(t, \alpha )>$.  
With increasing resistance we have thus a {\it cross over} from a state 
with a well defined persistent current (small mean-squared fluctuations)
to a fluctuation dominated state in which the mean-squared fluctuations 
of the persistent current are much larger than the average 
persistent current. For the derivation of these results we refer the reader to 
Refs. \cite{cedr} and \cite{annals}.

Here we pursue a discussion 
based on Langevin equations \cite{lange} 
similar to the approach described above. 
This approach is valid only for weak coupling constants $\alpha \ll 1$
but has the benefit of being simple.

We want to find the time evolution of a state $\psi(t)$
of the two-level system in the presence of the bath. 
We write the state of the two-level system 
\begin{equation}
\label{state}
\psi = e^{i\chi/2}
{\cos{\theta \over 2} \, e^{i\varphi/2} \hfill
\choose
\sin{\theta \over 2} \, e^{-i\varphi/2}},
\end{equation}
with $\theta$, $\varphi$ and $\chi$ real.  This is the most general
form of a normalized complex vector in two dimensions.  In terms of
$\theta$, $\varphi$ and the global phase $\chi$, the time dependent
Schr\"odinger equation reads \cite{lange} 
\begin{eqnarray}
\label{central1}
\dot{\varphi} &=& -\varepsilon_{0} - \delta \varepsilon (t) 
- \Delta_0 \cot\theta \cos\varphi, \\
\label{central2}
\dot{\theta} &=& -\Delta_0 \sin\varphi, \\
\label{chideq}
\dot{\chi} &=&  \Delta_0 {\cos\varphi \over \sin\theta}.
\end{eqnarray}
As shown by Eq.~(\ref{chideq})
the phase $\chi$ is
completely determined by the dynamics of the phases $\theta$ 
and $\varphi$ and has no back-effect 
on the evolution of $\theta$ and  $\varphi$.
While $\chi$ is irrelevant for expectation
values, like the persistent current or the charge on the dot,  it
plays an important role, in the discussion of phase diffusion times.

To close the system of equations 
we have now to find the voltage which drops across the system. 
In contrast to the two-level system discussed above in which the charge 
of the system is fixed, in the two-level system considered here 
the charge is permitted to tunnel 
between the dot and the arm of the ring. 
This charge transfer permits an additional displacement current 
through the system which we have to include to find the voltage
fluctuations.

The charge operator on the dot for our effective two-level problem is 
\begin{equation}
\hat{Q}_d = e \left( \matrix{ 1 & 0 \cr 0 & 0 } \right) . 
\end{equation}
Its quantum mechanical expectation value is 
\begin{equation}
{Q}_d = \langle \psi(t) | \hat{Q}_d
| \psi(t) \rangle = e \, \cos^{2}{\theta \over 2} .
\end{equation}
The displacement current is proportional to the time-derivative of 
this charge, 
$ \dot{Q}_d = -{e \over 2} \sin\theta \, \dot{\theta}$
multiplied by a ratio of capacitances which has to 
be found from circuit analysis. 
For this analysis we refer the reader to Ref. \cite{lange}.
We find that the  total current through the system is now 
given by $C_0  \dot{V} -
({C_0 /C_i}) ({e /2}) \sin\theta \, \dot{\theta}$. 
Using this result we find from the conservation of 
all currents (current through the system, the resistor and current of 
the noise source) that the fluctuating voltage 
across the system is determined by \cite{lange} 
\begin{equation}
\label{central3}
V = - C_0 R \,  \dot{V} -
R \, {C_0 \over C_i} \, {e \over 2} \, \sin\theta \, \dot{\theta}
+ R I_{N}(t) .
\end{equation}
Eqs.~(\ref{central1},~\ref{central2}) and
(\ref{central3}) form a closed system of equations in which the
external circuit is incorporated in terms of a fluctuating current
$I_{N}(t)$ and of an ohmic resistor $R$. In the next section, we
investigate Eqs.~(\ref{central1},~\ref{central2}) and
(\ref{central3}) to find the effect of zero-point fluctuations on
the persistent current of the ring.

\section{Fluctuations of the ground state}
\label{expansion}

First, let us discuss the stationary states of the system of
differential equations, Eqs.~(\ref{central1},
\ref{central2}) and (\ref{central3}) in the absence of the
noise term $I_{N}(t)$. We have  
$\sin\varphi = 0$ and
consequently a stationary state has $\varphi \equiv \varphi_0$, with
$\varphi_0 = 0$ or $\varphi_0 = \pi$. With this it is easy to show
that in the stationary state we must have $\theta \equiv \theta_0$,
with
\begin{equation}
\cot\theta_0 = \pm {\varepsilon_0 \over \Delta_0}. 
\end{equation}
The lower sign applies for $\varphi_0 = 0$.  This is the {\it ground state}
for the ring-dot system at fixed $\varepsilon(t) \equiv
\varepsilon_0$, and the upper sign holds for $\varphi_0 = \pi$.
The energy of the ground state is $-\hbar\Omega_0/2$, thus the global
phase is $\chi_0(t) = \Omega_0 t$ where $\Omega_0$
is the resonance frequency of the (decoupled) two-level system
(see Eq. (\ref{freee})).  
We also introduce 
the ``classical'' relaxation time $\tau_{RC} \equiv RC_0$, and a
rate
\begin{equation}
\label{Gamma}
\Gamma \equiv \pi \alpha {\Delta_0^2 \over \Omega_0}
\end{equation}
which as 
we shall see
is a relaxation rate due to the coupling of the ring-dot system 
to the external circuit. $\alpha$ is the dimensionless 
coupling constant introduced above (see Eq. (\ref{coupp})). 

Now, we switch on the noise $I_{N}(t)$.  We seek $\varphi(t) ,
\theta(t)$, $\chi(t)$ and $V(t)$ to linear order in the noise
current $I_{N}(t)$.  We expand $\varphi(t)$ and $\theta(t)$ to
first order around the ground state, $\varphi = 0$ and $\theta =
\theta_0$. For $ \delta \varphi(t) = \varphi(t) - \varphi_0$,
$\delta\theta(t) = \theta (t) - \theta_0$, we find in Fourier space,
\begin{eqnarray}
\label{central1linF}
-i \omega \delta\varphi &=& -\delta\varepsilon
+ {\Omega_0^2 \over \Delta_0} \, \delta\theta, \\
\label{central2linF}
-i \omega \delta\theta &=& -\Delta_0 \, \delta\varphi, \\
\label{central3linF}
-i \omega \, \delta \varepsilon
&=& {1 \over \tau_{RC}} \left[
-\delta\varepsilon - \Gamma \, \delta\varphi
+ {e \over \hbar} \, R \, {C_0 \over C_i} \, I_{N}
\right]. 
\end{eqnarray}
We also expand the global phase $\chi(t)$ around its
evolution in the ground state $\chi_0(t) = \Omega_0 t$, and define
$\delta\chi(t) = \chi(t) - \chi_0(t)$.  In Fourier space,
Eq.~(\ref{chideq}) becomes
\begin{equation}
\label{chideqF}
-i\omega \delta\chi = \Omega_0 \, {\varepsilon_0 \over \Delta_0}
\delta\theta.
\end{equation}
We note that there is no effect of the global shift in energy,
$\hbar\nu(t)$, as it is quadratic in the voltage $\delta V$, and we
are only interested in effects up to linear order in $\delta V$.

\section{Mapping onto a harmonic oscillator} 

Let us assume that the charge relaxation time of 
the external circuit $\tau_{RC}$ is very short compared to the dynamics 
of the two-level system $\tau_{RC} \ll \Omega_{0}$.
Eliminating $\delta\theta$ with the help 
of Eq. (\ref{central2linF}) and  $\delta\varepsilon$
with the help of Eq. (\ref{central3linF}) we find 
\begin{eqnarray}
\label{oscillator}
(\omega^{2} - i \omega  \Gamma - \Omega_0^2 ) \, \delta\theta
= \Delta_0 \, {e \over \hbar} \, {C_0 \over C_i} \, R \, I_{N} .
\end{eqnarray}
Thus we have mapped the dynamics of the fluctuations 
away from the ground state of this two-level system 
on the Langevin equation of a damped harmonic oscillator
subject to quantum fluctuations. 
$\delta\theta$ plays the role of the charge,  
$\delta\varphi$ the role of the current 
and $\Gamma$ takes the role of the friction constant
in the $LC$-oscillator discussed in Section \ref{lamb}. 
We can now immediately use the results of the first few sections 
of this work to describe the fluctuations in the ground state 
of this two-level problem.
The spectral density $S_{\theta\theta}(\omega)$
is just that of the coordinate of the harmonic oscillator 
or that of the charge of the $LC$-oscillator,  
\begin{equation}
S_{\theta\theta} (\omega)
= {2\pi \, \alpha \, \Delta_0^2 \, | \omega |
\over
\left[
\left( \omega^2 - \Omega_0^2 \right)^2
+ \Gamma^2 \omega^2 
\right]}.
\end{equation}
Note that the intensity of the noise power is proportional to $\Delta_0^2$.
Alternatively we could write the intensity as 
$\pi \alpha \Delta_0^2 = \Omega_0 \Gamma$. 

The approach presented here is valid only in the weak 
coupling limit. The poles of the weakly damped oscillator 
are 
\begin{equation}
\omega_\pm = \pm \Omega_0 - i{\Gamma \over 2},
\end{equation}
and we can write 
\begin{equation}
\label{thetasp1}
S_{\theta\theta} (\omega)
= {2\pi \, \alpha \, \Delta_0^2 | \, \omega |
\over
\left[
\left( \omega - \Omega_0 \right)^2
+ \left( \Gamma / 2 \right)^2
\right]
\left[
\left( \omega + \Omega_0 \right)^2
+ \left( \Gamma / 2 \right)^2 
\right]}. 
\end{equation} 
Expressing the spectrum as a sum of separate pole contributions 
we find 
\begin{equation}
\label{thetasp2}
S_{\theta\theta} (\omega)
= \frac{2\pi \, \alpha \, \Delta_0^2}{\Omega_{0}}
\left( \frac{1 - ((\omega - \Omega_0 )/4 \Omega_0 )}
{\left[ \left( \omega - \Omega_0 \right)^2 + 
\left( \Gamma / 2 \right)^2 \right]}
+ 
\frac{1 + ((\omega + \Omega_0 )/4 \Omega_0 )}
{\left[\left( \omega + \Omega_0 \right)^2
+ \left( \Gamma / 2 \right)^2
\right]} \right). 
\end{equation}
The factors ${1 \mp (\omega \mp \Omega_0 )/4 \Omega_0 }$
account for the effect of the Lorentzian tails of the far away pole. 

In the literature \cite{ww,gsw,costi} it is often 
the correlation function of $\sigma_z$ which is of interest. 
We have $<\psi | \sigma_z |\psi> = \cos(\theta)$
and thus for the fluctuations away from the average 
$\Delta <\psi | \sigma_z |\psi> = - \sin(\theta_{0}) \delta \theta$.
Since $\sin(\theta_{0}) = \Delta_{0}/\Omega_{0}$, 
we find in the zero-temperature limit
$S_{\sigma_{z}\sigma_{z}} (\omega) = (\Delta^{2}_{0}/\Omega^{2}_{0})
S_{\theta\theta} (\omega)$. 
This result agrees with an expression given by 
Weiss and Wollensak \cite{ww} and G\"orlich et al. \cite{gsw}
who have used an entirely different approach. 
For non-zero temperatures Weiss and Wollensak find in addition a 
peak around zero-frequency: This is a Debye relaxation peak 
and in the discussion given here it is not included. We have  
expanded around the (time-independent) ground state 
of the decoupled system. To find the Debye relaxation 
peak from a weak coupling treatment it is necessary to consider 
the relaxation towards the instantaneous (time-dependent) 
state of the system. 
At temperatures $kT \ll \hbar \Omega_{0}$ the weight of 
the Drude peak is exponentially small. The result of 
Weiss and Wollensak also includes a temperature dependent
renormalization of the bare pole frequencies $\Omega_{0}$. 
The most essential 
point for the discussion here, is however, the fact that the 
peaks are broadened with a relaxation rate $\Gamma$. The spectrum 
Eq. (\ref{thetasp2}) is that of the coordinate (or charge)
of a weakly damped harmonic oscillator 
subject to quantum fluctuations.

Using Eqs. (\ref{central2linF}) and (\ref{central3linF})
we obtain similarly the spectral densities $S_{\varphi \varphi}(\omega)$, 
$S_{\chi\chi}(\omega)$ and cross-correlations like 
$S_{\theta \varphi}(\omega)$.

\section{Suppression of the Persistent current}
\label{pc}
Let us next examine the reduction of the persistent current 
using the approach outlined above. 
To be brief we consider only the case 
of a symmetric ring $t_{R} = t_{L} \equiv t$ 
and $C_R = C_ L$ (see Fig. \ref{system}). 
The persistent current is the quantum and statistical 
average of the operator 
\begin{equation}
\hat{I}_c = {\cal J} \sigma_z 
\label{icop}
\end{equation}
where $\cal J$ is given by
\begin{equation}
{\cal J}
= {\hbar c} {\partial \Delta_0 \over \partial \Phi}.
\end{equation}
In general, in the non-symmetric
case, the operator for the persistent
current depends also on the capacitances
(see Appendix B of Ref. \cite{lange}). 

The quantum mechanical 
expectation value of the persistent current for the state given in
Eq.~(\ref{state}) reads
\begin{equation}
I(t) \equiv \langle \psi(t) | \hat{I}_c | \psi(t) \rangle
= {1 \over 2} {\rm Re}\, ( {\cal J} \sin\theta \, e^{-i\varphi}).
\end{equation}
We are interested in the {\em statistically averaged} 
persistent current $\langle I(t) \rangle$.  
Therefore, we have to calculate the correlator $\langle \sin\theta \,
e^{-i\varphi} \rangle$.  First, we observe that there are no
correlations between $\delta\varphi$ and $\delta\theta$. 
The symmetrized correlation function  
$\langle \delta\varphi(t) \delta\theta(t) \rangle = 0$,
since the spectral density $S_{\varphi\theta}(\omega)$ is an odd function
of $\omega$. (This statement is equivalent to the vanishing of 
the correlations between $p$ and $q$ for a harmonic oscillator). 
For a harmonic oscillator
the fluctuations are Gausssian 
and thus within the range of validity of our discussion there are 
no correlations to all orders in $\theta (t)$ and 
and $\varphi(t)$. Thus we have 
$ \langle \sin\theta e^{-i\varphi} \rangle
= \langle \sin\theta \rangle
\langle e^{-i\varphi} \rangle $. 
and 
\begin{equation}
\label{dphi} 
\langle e^{-i\varphi(t)} \rangle
= e^{-i\varphi_0} \langle e^{-i\delta\varphi(t)} \rangle
= \exp\left( -{\left\langle\delta\varphi^2(t) \right\rangle \over 2} \right),
\end{equation}
where we have used that $\varphi_0 = 0$.  In the weak coupling limit,
and in the extreme quantum limit,
$T=0$, we find for the time averaged mean-squared fluctuations
to leading order in $\Gamma$, 
\begin{equation}
\label{msdphi2}
\langle \delta\varphi^2(t) \rangle
= \int_0^{\omega_c} \frac{d\omega}{\pi} 
S_{\varphi\varphi}(\omega)
= \frac{\Omega_0}{\Delta_0^{2}} 
[ 2 \Gamma \ln {\omega_c \over \Omega_0} - \Gamma + \pi \Omega_{0} ]
\approx 2\alpha
\ln {\omega_c \over \Omega_0} . 
\end{equation}
Here we have assumed that the cut-off frequency is so large 
that the logarithmic term dominates. 
In the limit $\omega_c \gg \Omega_0$, we
can neglect $\langle \delta\theta^2(t) \rangle = \int_0^{\omega_c}
(d\omega/\pi) S_{\theta\theta}(\omega)$ against $\langle
\delta\varphi^2(t) \rangle$.  We insert $\langle \delta\varphi^2(t)
\rangle$ and $\sin\theta_0 = \Delta_0/\Omega_0$ into $\langle
\sin\theta \, e^{-i\varphi} \rangle$, and find 
a noise averaged persistent current in the ring given by 
\begin{equation}
\label{pcsimple}
\langle I(t) \rangle
= -{\hbar c \over 2} {\partial \Delta_0 \over \partial \Phi}
{\Delta_0 \over \Omega_0}
\left( {\Omega_0 \over \omega_c} \right)^\alpha.
\end{equation}
The weak coupling limit corresponds to 
$\alpha \ll 1$. 
The power law for the persistent current obtained in
Eq.~(\ref{pcsimple}), as well as the exponent $\alpha$,
Eq.~(\ref{coupp}) coincide in this limit 
with the result obtained by using a Bethe
ansatz solution.  
Cedraschi et al. \cite{cedr} 
found for $\alpha < 1$
that at resonance ($ \epsilon = 0 $) the persistent current 
is given by 
$ I(\varepsilon=0) \propto
\left(
{\Delta_0 /\omega_c}
\right)^{\alpha / (1-\alpha )}.
$
For a small coupling parameter, $\alpha \ll 1$, the Bethe ansatz
result goes over to the power law of
Eq.~(\ref{pcsimple}).  Thus, if it can be assumed that
the logarithmic term dominates in Eq. (\ref{msdphi2}), 
the simplified discussion presented
here leads, at least in the weak coupling limit, 
to the same result as
the one obtained in Ref.~\cite{cedr}.

We emphasize that the persistent current is a property of the ground
state of a system. In our case, the persistent current is, however,
carried by only a part of the system.  Due to the coupling to the
external circuit this subsystem is subject to fluctuations which even
at zero temperature suppress the persistent current.  If we keep the
capacitances fixed, Eq.~(\ref{pcsimple}) 
gives a persistent current which decreases with increasing external
resistance $R$.  

For the case considered here it is simple to also discuss the 
instantaneous fluctuations of the persistent current. 
From Eq. (\ref{icop}) we find $\hat{I}_{c}^{2} = {\cal J}^{2} {\bf 1}$
where ${\bf 1}$ is the unit matrix. Thus the mean squred fluctuations
$(\Delta I )^{2} = (\hat{I}_{c} - \langle I (t) \rangle )^{2}$
of the persistent current are 
\begin{equation}
\langle (\Delta I )^{2} \rangle = {\cal J}^{2} - 
( \langle I(t) \rangle )^{2}
\end{equation}
Thus with increasing coupling constant $\alpha$ 
the average persistent current decreases and the mean squared fluctuations 
of the persistent current increase (see also Fig. \ref{kondopc}). 

We next characterize the fluctuations of the
angle variables of the ground state in more detail.

\section{Phase Diffusion Times}
\label{dephasing}

Due to the vacuum fluctuations the ground state of the 
two-level system with tunneling is a dynamic state. 
To see this we project the actual state of the 
system on the 
{\em ground state\/} $\psi_- = (\cos\theta_0/2, \sin\theta_0/2)$, with
energy eigenvalue $-\hbar\Omega_0/2$, and the {\em excited state\/} $\psi_+ =
(-\sin\theta_0/2, \cos\theta_0/2)$ with eigenvalue $\hbar\Omega_0/2$.
Instead of the wave function $\psi(t)$ it is more convenient 
to consider $\psi_R(t) \equiv
\exp(i\hat{H}_0 t/\hbar) \psi(t)$. 
To first order in $\delta\varepsilon$, we find for the wave function 
$\psi_R$, 
\begin{equation}
\psi_R(t) = ( 1 + i \alpha ) \psi_{-} + \beta \psi_{+} e^{i\Omega_0 t} .
\end{equation}
with 
\begin{equation}
\alpha = {\delta\chi(t) \over 2}
- {\varepsilon_0 \over \Omega_0} {\delta\varphi(t) \over 2} \, 
\end{equation}
\begin{equation}
\beta = \left[ {\delta\theta(t) \over 2}
- i{\Delta_0 \over \Omega_0} {\delta\varphi(t) \over 2} 
\right] e^{i\Omega_0 t} .
\end{equation}
It is useful to write $\alpha (t)$ and  $\beta (t)$ 
in terms of the Fourier amplitudes of the fluctuating 
phase $\phi (t)$. 
We find 
\begin{equation}
\alpha (t) = {1 \over 2} {\varepsilon_0 \over \Omega_0}
\int {d\omega \over \omega^{2}} 
( \Omega_0^{2} - \omega^{2} ) \phi_{\omega} \exp(-i\omega t) , 
\end{equation}
\begin{equation}
\beta (t) = {1 \over 2} {\Delta_0 \over \Omega_0}
\int {d\omega \over i\omega} (\Omega_0  +  \omega ) 
\phi_{\omega} \exp(-i(\omega - \Omega_0)t) . 
\end{equation}
The term $\exp(i\Omega_0 t)$ arises due to the energy 
difference between the ground state and the excited state.   
Thus the projection amplitudes 
$<\psi (t)|\psi_{-} (t)> = 1+ i \alpha (t)$
and $<\psi (t)|\psi_+ (t)> = \beta (t)$
exhibit in time a mean-squared deviation away from their 
initial value given by 
\begin{eqnarray}
\label{cminus}
\left\langle \left| \alpha (t) - \alpha (0) \right|^2 \right\rangle 
= {\varepsilon_0^2 \over \Omega_0^2}
\int {d\omega \over 2\pi \omega^{4}} \sin^2 {\omega t\over 2}
( \Omega_0^{2} - \omega^{2} )^{2} S_{\varphi\varphi}(\omega), \\
\label{cplus}
\left\langle \left| \beta (t) - \beta (0) \right|^2 \right\rangle 
\approx {\Delta_0^2 \over \Omega_0^2}
\int {d\omega \over 2\pi } \sin^2 {\omega t \over 2}
S_{\varphi\varphi}(\omega+\Omega_0)
\end{eqnarray}
\noindent
The long time behavior of Eq.~(\ref{cminus}) is dominated by the
frequencies near $\omega=0$.  The spectral density
$S_{\varphi\varphi}(\omega)$ vanishes like $\omega^2$ for finite
temperatures or even like $|\omega|^3$ in the zero-temperature limit.
For finite temperatures this gives raise to a long time
behavior of the type $\langle |\alpha(t) - \alpha (0)|^2 \rangle \sim
t/\tau_-$, with a characteristic phase diffusion time 
\begin{equation}
\label{tauminus}
\tau_- = {\hbar \over 2\pi\alpha \, kT}
{\Omega_0^2 \over \varepsilon_0^2}.
\end{equation}
Note that this behavior arises from the fluctuations in the global 
phase $\chi$. Note also that $\tau_-$ 
depends on the detuning 
$\varepsilon_0$. In particular,
at resonance $\varepsilon_0=0$, the phase diffusion time $\tau_-$
diverges for any temperature.

The long time behavior of Eq.~(\ref{cplus}), on the other hand,
is determined by the frequencies near $\Omega_0$.  In the
vicinity of this characteristic frequency, $S_{\varphi\varphi}
(\Omega_0+\omega)$ 
shows a $\omega^{-2}$ behavior {\em at finite as well as at zero
temperature}, which is cut off by the relaxation rate $\Gamma$,
defined in Eq.~(\ref{Gamma}) at very small frequencies $\omega
\sim \Gamma \ll \Omega_0$.  We have, for $|\omega| \ll
\Omega_0$
\begin{equation}
S_{\varphi\varphi}(\Omega_0+\omega)
\approx
{2\pi\alpha \, \Omega_0 \coth {\hbar\Omega_0 \over 2kT}
\over \omega^2 + (\Gamma/2)^2}. 
\end{equation}
The time evolution
of Eq.~(\ref{cplus}) for times much larger than the inverse of
the characteristic frequency $\Omega_0$, yet smaller than the inverse
of the relaxation rate $\Gamma$, is therefore linear in time with a
characteristic time $\tau_+$, where
\begin{equation}
\label{tauplus}
\tau_+ = {1 \over \Gamma}
\tanh {\hbar\Omega_0 \over 2kT}.
\end{equation}
Note that Eq.~(\ref{tauplus}) holds for finite temperatures as
well as in the quantum limit.  The phase diffusion time $\tau_+$ is
inversely proportional to $T$ at high temperatures,
\begin{equation}
\label{taupf}
\tau_+ = {1 \over \Gamma} {\hbar\Omega_0 \over 2kT}, \quad
(kT \gg \hbar\Omega_0),
\end{equation}
just as the other characteristic time $\tau_-$.  In the low-temperature or
quantum limit, however, it {\em saturates\/} to a value
\begin{equation}
\label{taupq}
\tau_{+} = {1 \over \Gamma}, \quad (kT \ll \hbar\Omega_0).
\end{equation}
The crossover from high
temperature behavior to the quantum limit behavior takes place at $kT
\sim \hbar\Omega_0$.

We emphasize that Eq.~(\ref{cminus}) and Eq.~(\ref{cplus})
do not hold for arbitrarily long times.  In reality the mean-square
displacements $\langle |\alpha (t) - \alpha (0)|^2 \rangle$ 
and $\langle |\beta (t) - \beta (0)|^2 \rangle$ are bounded
since the wave function $\psi_R(t)$ is normalized to 1.  The fact that
Eq.~(\ref{cminus}) grows
without bounds is an artifact of the linearization of
Eqs.~(\ref{central1})--(\ref{chideq}) and
Eq.~(\ref{central3}).
The phase-diffusion rates $\tau_-$ and $\tau_+$ 
can be related to the relaxation rate and the dephasing time 
given by Weiss and Wollensak \cite{ww} and 
Grifoni, Paladino and Weiss \cite{grif}. 
To find the dephasing rate we write 
the time evolution of the state $\Psi_{R}(t)$ (see Eq. (54)) with the help 
of an overall phase $\eta(t)$ in the form $\Psi_{R}(t) = \exp(-i\eta(t))
\Psi_{R}(0)$. For times scales over which $\eta$ remains small we have 
$\Psi_{R}(t) = (1-i\eta(t)) \Psi_{R}(0)$ or 
$\Psi_{R}(t) - \Psi_{R}(0) = -i\eta(t) \Psi_{R}(0)$. 
The scalar product of $\Psi_{R}(t) - \Psi_{R}(0)$ with itself 
is $|\Psi_{R}(t) - \Psi_{R}(0)|^{2} = |\eta(t)|^{2}$ 
and its expectation value 
is thus just $<|\eta(t)|^{2}> = <|c_{-}(t) - c_{-}(0)|^{2} > + 
<|c_{+}(t) - c_{+}(0)|^{2}>$. The dephasing rate is 
$\Gamma_{\phi} = <|\eta(t)|^{2}>/2t$ and is therefore given by 
\begin{equation}
\label{dephtime} \Gamma_{\phi}  = 
\frac{1}{2\tau_+} + \frac{1}{2\tau_-} =  
{\Gamma \over 2} \coth {\hbar\Omega_0 \over 2kT} +
{\pi\alpha \, kT \over \hbar}
{\varepsilon_0^2 \over \Omega_0^2}.
\end{equation}
This dephasing rate, valid on short and intermediate times, 
saturates in the zero-temperature limit. At $kT = 0$ we have 
$\Gamma_{\phi} = \Gamma /2$.  

To summarize, we find that in a two-level system
with tunneling, for which the Hamiltonian does not commute 
with the total Hamiltonian, phase-diffusion times 
$\tau_-$ and $\tau_+$, which are related to the projection of 
the equilibrium state
$\psi(t)$ onto the ground state and the excited state. 
These rates also determine the dephasing rate. 
If we represent the state of the system as a spin, these results 
show that even in the ground state, the spin undergoes 
diffusion around the point on the Bloch 
sphere which it would mark in the absence of the bath. 
We have thus a problem for which the ground state 
exhibits only partial coherence. 

\section{Energy Fluctuations}

It is interesting to compare the energy fluctuations 
of the two-level system with those of the harmonic 
oscillator in the Lamb model. Using the Bloch state vector 
Eq. (\ref{state}) and the operator Eq. (\ref{ham2})
for the system we find for the average energy 
\begin{equation}
\label{ham3}
\langle \Psi |H_{0}|\Psi \rangle = 
(\hbar \epsilon_0 /2) cos (\theta) 
- (\hbar \Delta_0 /2) sin(\theta) cos(\phi) .
\end{equation}
For the ground state in the absence of the bath we have 
$\theta = \theta_{0}$ determined by $sin(\theta_{0}) = \Delta_{0}/\Omega_{0}$
and $\phi = \phi_{0} = 0$ and we find immediately, 
$\langle \Psi |H_{0}|\Psi \rangle = - \hbar \Omega_{0} /2$. 
If we know couple the system to the bath the angle variables fluctuate
away from these values. To find the effect of these fluctuations 
we replace $\theta$ and $\phi$ in Eq. (\ref{ham3}) by
$\theta = \theta_{0} + \theta - \theta_0$
and $\phi = \phi_{0} + \phi - \phi_0$. 
We proceed as in the discussion of the persistent current
and consider the fluctuations in $\theta$ as small compared 
to the fluctuations in $\phi$. Using 
$\langle cos(\theta - \theta_0 ) \rangle > \approx 1$ 
and taking into account that $\phi_0 = 0$, we obtain, 
\begin{equation}
\label{ham4}
\langle \Psi |H_{0}|\Psi \rangle = 
(\hbar \epsilon_0 /2) cos(\theta_0 )
- (\hbar \Delta_0 /2) sin(\theta_0 ) \langle cos(\phi- \phi_0 ) \rangle .
\end{equation}
Using Eqs. (\ref{dphi}) and  (\ref{msdphi2}), assuming that the logarithmic 
term is dominant, we obtain, 
\begin{equation}
\label{ham5}
\langle \Psi |H_{0}|\Psi \rangle = - \hbar \Omega_{0} /2 
+ (\hbar \Delta_0^{2} /2 \Omega_{0}) (1 - (\Omega_{0}/\omega_c )^{\alpha}) .
\end{equation}
It is now easy to find the fluctuations 
in the energy of the two-level system. 
Using the expression of the energy operator Eq. (\ref{ham3}) 
we find, 
\begin{equation}
H_{0}^{2} = \frac{\hbar^{2} \Omega_{0}^{2}}{4} {\bf 1} 
\end{equation}
where ${\bf 1}$ is the unit matrix. Therfore,  
to leading order in the coupling constant,
we find  
\begin{equation} 
\langle (\Delta E)^{2} \rangle = 
\langle \Psi |H_{0}^{2}|\Psi \rangle 
- \langle \Psi |H_{0}|\Psi \rangle^{2}  = 
\frac{\hbar^{2} \Delta_0^{2}}{4} 
\left( 1 - \left(\frac{\Omega_{0}}{\omega_{c}} \right)^{\alpha} \right) 
\end{equation}
Thus the energy of a two-level system coupled to a bath 
fluctuates. The fluctuations increase rapidly 
with increasing coupling constant.  

Measurements of 
energy fluctuations are possible: for instance (as done in optics)
by resonantly exciting the system from one of the two levels 
to a still higher third level \cite{optics}. 
Clearly it would be very intersting 
to see such an experiment performed for two level systems in a 
mesoscopic system.

\section{Discussion}
\label{discussion}

In this work we have investigated the fluctuations 
of the ground state of a system coupled to a bath. 
We have emphasized that due to coupling 
to vacuum fluctuations the energy of a system is not sharp 
but fluctuates in time \cite{cedr}. We have demonstrated this 
with an explicite calculation for an $LC$-oscillator 
coupled to a transmission line.  
Such energy fluctuations are not a consequence 
of absorption or emission of photons (real transitions) 
but simply reflect the fact that 
a normal mode of the uncoupled system is not a normal mode 
of the system coupled to the bath. Any appeal to purely statistical 
mechanics arguments which treats the system and bath modes as if 
they were uncoupled (neglects the coupling energy) 
simply misses this phenomenon. We have examined 
the trace which the system leaves in the environment.
We have found that this trace has its origin in the correlations 
between the incident and the out-going vacuum field fluctuations. 

We emphasize that the system-bath interaction is 
treated here in a non-perturbative way: even in the weak 
coupling limit the spectral densities which characterize the fluctuations 
of the angle variables of the ground state depend to all 
orders on the coupling constant. This is obviously very different 
form a perturbative treatment. Perhaps equally important 
is the following: Our model as an electronic interpretation: 
yet we do not start by considering, say the interaction of a single 
electron with a two-level fluctuator. Instead, as the transmission 
line illustrates, what is considered is the interaction of plasmon waves 
(collective excitations) with the two-level system. 

We have investigated the decoherence of a two-level system 
for which the Hamiltonian of the system commutes 
with the total Hamiltonian. This in an example of a system 
in which the ground state remains a pure state
even in the presence of the bath. Superpositions of the ground state and 
the excited state decoher 
due to vacuum fluctuations without energy transfer \cite{ekert}
although only inversely proportional to a power law in time. 
The main focus in this work has been on a simple model 
system with tunneling for which the Hamiltonian of the system 
does not commute with the total Hamiltonian. The persistent current 
provides a measure of the coherence of the ground state and 
is suppressed with increasing coupling to the bath.
In the weak coupling limit,   
in the ground state, the system exhibits a spectral density 
for the fluctuations of the angles of the spin on the Bloch 
sphere which is that of a damped harmonic 
oscillator subject to vacuum fluctuations. This enabled us 
to show that on short and intermediate 
times this system undergoes diffusion 
on the Bloch sphere even in the ground state. 
We have thus a system with a ground state 
that is only partially coherent. 

What are the implications of these results for the 
discussion of the dephasing times in mesoscopic physics? 
The fact that the commutation of the systems Hamiltonian 
with the total Hamiltonian determines in the above mentioned 
examples whether or not we have a coherent ground state 
is possibly a general rule with which 
we can decide whether to expect a dephasing rate 
which tends to zero with temperature or a dephasing rate 
which saturates. In weak localization an electron and its time-reversed 
companion are the "system" and all the other electrons,  
together with the electromagnetic interaction,  
provide the bath. Does the Hamiltonian of the quasi particles 
commute with the total Hamiltonian? Since weak localization
also invokes an ensemble averaging procedure the answer 
to this question is not obvious.

Vacuum fluctuations have played a key role in the development 
of quantum mechanics. Our increasing ability to make small systems 
and measure them makes it likely that these fluctuations 
will continue to be of high interest. 

\section*{Acknowledgement}
\label{acknowledgement}

It is a pleasure to thank Kirill Nagaev who has collaborated with me 
on energy fluctuations generated by the vacuum. 
I thank Georg Seelig for help with the figures 
and Harry Thomas, Daniel Loss and Daniel Braun for discussions.


\begin{thebibliography}{99}

\bibitem{mill}     P. W. Milloni, {\it The Quantum Vacuum}, 
                   (Academic Press, Boston, 1994). 



                   
\bibitem{legg}     A. J. Leggett, S. Chakravarty, A. T. Dorsey, 
                   M. P. A. Fisher and W. Zwerger, 
                   Rev. Mod. Phys. {\bf 59}, 1 (1987).                    

\bibitem{weiss}    U. Weiss, 
                   {\it Quantum Dissipative Systems}, (Word Scientific, 2000).
                   
\bibitem{aleiner2} I.~L. Aleiner, B.~L. Altshuler, and M.~E. Gershenson, 
                   Waves in Random Media {\bf 9}, 201 (1999).                    

\bibitem{ekert}    G. M. Palma, K.-A. Souminen, and A. Ekert,
                   Proc. Royal Soc. London, A {\bf 452}, 567 (1996). 




\bibitem{mohanty1} P. Mohanty, E.~M.~Q. Jariwala, and R.~A. Webb, 
                   Phys. Rev. Lett. {\bf 78},  3366 (1997).      
      
\bibitem{mohanty2} P. Mohanty, 
                   Ann. Physik {\bf 8}, 549 (1999).      
      
\bibitem{golubev}  D.~S. Golubev and A.~D. Zaikin, 
                   Phys. Rev. Lett. {\bf 82},  3191  (1999);
                   A. D. Zaikin and D. S. Golubev, 
                   Physica B {\bf 280}, 453 (2000). 

\bibitem{aleiner1} I.~L. Aleiner, B.~L. Altshuler, and M.~E. Gershenson, 
                   Phys. Rev. Lett. {\bf 82},  3190  (1999).       

\bibitem{natel}    D. Natelson, R. L. Willett, K. W. West, 
                   and L. N. Pfeiffer,
                   Phys. Rev. Lett. {\bf 78}, 1821 (2001).  
                   A large set of closely related experiments is reported 
                   by J. J. Lin and L. Y. Kao, J. Phys. Condens. Matter
                   {\bf 13}, L119 (2001).              

\bibitem{mbcs}     M. B\"uttiker and C.~A. Stafford, 
                   Phys. Rev. Lett. {\bf 76},  495  (1996).
                   
\bibitem{cedr}     P. Cedraschi, V.~V. Ponomarenko, and M. B\"uttiker, 
                   Phys. Rev. Lett. {\bf 84}, 346  (2000).


\bibitem{chak}     S. Chakravarty, Phys. Rev. Lett. {\bf 49}, 681 (1982).

\bibitem{bray}     A. J. Bray and M. A. Moore, Phys. Rev. Lett. {\bf 49}, 
                   1545 (1982). 

\bibitem{loss}     D. Loss and K. Mullen
                   Phys. Rev. B {\bf 43}, 13252  (1991).  
                                   
                                                    
\bibitem{lamb}     H. Lamb, 
                   Proc. London Math. Soc. {\bf 53}, 208 (1900).

\bibitem{ford}     G. W. Ford, J. T. Lewis, and R. F. O'Connel, 
                   Phys. Rev. A {\bf 37}, 4419 (1988).                    

\bibitem{yude}                   
                   B. Yurke and J. S. Denker, 
                   Phys. Rev. A {\bf 29}, 1419 (1984). 
                   
\bibitem{gard}     C. W. Gardiner and P. Zoller, {\it Quantum Noise}, 
                   (Springer, Heidelberg, 2001).              
                   

\bibitem{naga}     K. Nagaev and M. B\"uttiker, (unpublished). 

\bibitem{gavish}   U. Gavish, Y. Levinson, and Y. Imry, 
                   Phys. Rev. B {\bf  62}, R10637 (2000). 

\bibitem{becker}   R. Becker, 
                   {\it Theorie der W\"arme}, (Springer Verlag, Berlin, 
                   1966).
 

\bibitem{zure}     W. H. Zurek, 
                   Physics Today, October, 36 (1991);
                   
                   
\bibitem{sia}      A. Stern, Y. Aharonov, and Y. Imry, 
                   Phys. Rev. A {\bf 41}, 3436 (1990).  

\bibitem{annals}   P. Cedraschi and M. B\"uttiker,  
                   Annals of Physics, {\bf 289}, 1 - 23 (2001).

\bibitem{lange}    P. Cedraschi and M. B\"uttiker, 
                   Phys. Rev. B {\bf 63}, 165312 (2001).
                   
\bibitem{loma}     D. Loss and T. Martin                   
                   Phys. Rev. B {\bf 47}, 4619 (1993). 

\bibitem{bouch}    V. Bouchiat, D. Vion, P. Joyez, D. Esteve, 
                   and M. H. Devoret, 
                   J. of Superconductivity, {\bf 12}, 789 (1999). 
                   
                   
\bibitem{moji}     J. E. Moji et al, Science {\bf 285}, 1036 (1999). 
                   L. Tian, L.~S. Levitov, C.~H. van~der~Wal, J.~E. Mooij, 
                   T.~P. Orlando, S. Lloyd, C.~J.~P.~M. Harmans, 
                   and J.~J. Mazo, 
                   in "Quantum Mesoscopic Phenomena and Mesoscopic Devices
                   in Microelectronics", 
                   edited by I. O. Kulik and R. Ellialtioglu, 
                   (Kluwer, Netherlands, 2000). p. 429. 


\bibitem{makh}     Y. Makhlin, G. Sch\"on, and A. Shnirman, 
                   in  "Quantum Physics at Mesoscopic Scale"
                   edited by D.C. Glattli, M. Sanquer and 
                   J. Tran Thanh Van
                   (EDP Sciences, Les Ulis, 2000). p. 113  

\bibitem{bee}      C. W. J. Beenakker, Phys. Rev. B {\bf 44}, 1646 (1991).      

\bibitem{eckle}    H.-P. Eckle, H. Johannesson, and C.~A. Stafford, 
                   J. Low Temp. Phys. {\bf 118}, 475 (2000).  
                   cond-mat/0010101   

\bibitem{kang}     K. Kang and S.-C. Shin, Phys. Rev. Lett. {\bf 85}, 
                   5619 (2000).                   
                                    
\bibitem{affl}     I. Affleck and P. Simon, Phys. Rev. Lett. 
                   {\bf 86}, 2854 (2001); 
                   P. Simon, I. Affleck, cond-mat/0103175 .  
           
\bibitem{hu}       Hui Hu, Guang-Ming Zhang, Lu Yu, Phys. Rev. Lett. 
                   {\bf 86}, 5558 (2001). 
                

      



\bibitem{ww}       U. Weiss and M. Wollensak, 
                   Phys. Rev. Lett. {\bf 62}, 1663 (1989). 

\bibitem{gsw}      R. G\"orlich, M. Sassetti, and U. Weiss, 
                   Europhys. Lett. {\bf 10}, 507 (1989). 
                   
\bibitem{costi}    T. A. Costi and Kieffer, 
                   Phys. Rev. Lett. {\bf 76}, 1683 (1996).
                   

\bibitem{grif}     M. Grifoni, E. Paladino and U. Weiss, 
                   Eur. Phys. J. B{\bf 10}, 719 (1999). 
                   

\bibitem{optics}   A. Beige and G. C. Hegerfeldt, J. Mod. Optics, 
                   {\bf 44}, 345 (1997). 
                   
                   
\end{thebibliography}
\end{document}